\pdfoutput=1
\documentclass[nofootinbib,twocolumn,showpacs,amssymb,floatfix,deluxe,aps]{revtex4}

\usepackage [latin1]{inputenc}
\usepackage{graphicx}% Include figure files
\usepackage{dcolumn}% Align table columns on decimal point
\usepackage{bm}% bold math
\usepackage{epsfig}
\usepackage[english]{babel}
\usepackage{amsmath}
\usepackage{amssymb}

\usepackage{verbatim}
\usepackage{alltt}

\def\({\left(}
\def\){\right)}

\begin{document}

%\large{

%CERN-PH-TH/2010-120
\preprint{}
\title{Deciphering the AMS cosmic-ray positron flux} 
% Force line breaks with \\

\author{A. De R\'ujula${}^{a,b}$}
\affiliation{  \vspace{3mm}
${}^a$Instituto de F\'isica Te\'orica (UAM/CSIC), Univ. Aut\'onoma de Madrid, Spain;\\
${}^b$Theory Division, CERN, CH 1211 Geneva 23, Switzerland
}

\date{\today}% It is always \today, today,

\begin{abstract}
The flux of cosmic-ray high-energy positrons has recently been measured by AMS 
with unprecedented precision. This flux is well above the expectation from secondary positrons 
made by the observed fluxes of nuclear cosmic rays impinging on the interstellar medium.
 Various authors  
have pointed out that the positron excess may originate at the primary
cosmic-ray source itself, rather than in the more local ISM, thus avoiding the temptation to invoke a dark-matter decay or annihilation origin, or nearby pulsars. We investigate the possibility that the source is 
the one of a comprehensive model of gamma-ray bursts and cosmic rays, proposed two decades ago. The result, based on the original unmodified priors of the model --and with no fitting of parameters-- very closely reproduces the shape and magnitude of the AMS observations. 

\end{abstract}


\maketitle

\section{Introduction and outlook}

\begin{figure}[]
\vspace{-.5cm}
\hspace{-.5cm}
\centering
\epsfig{file=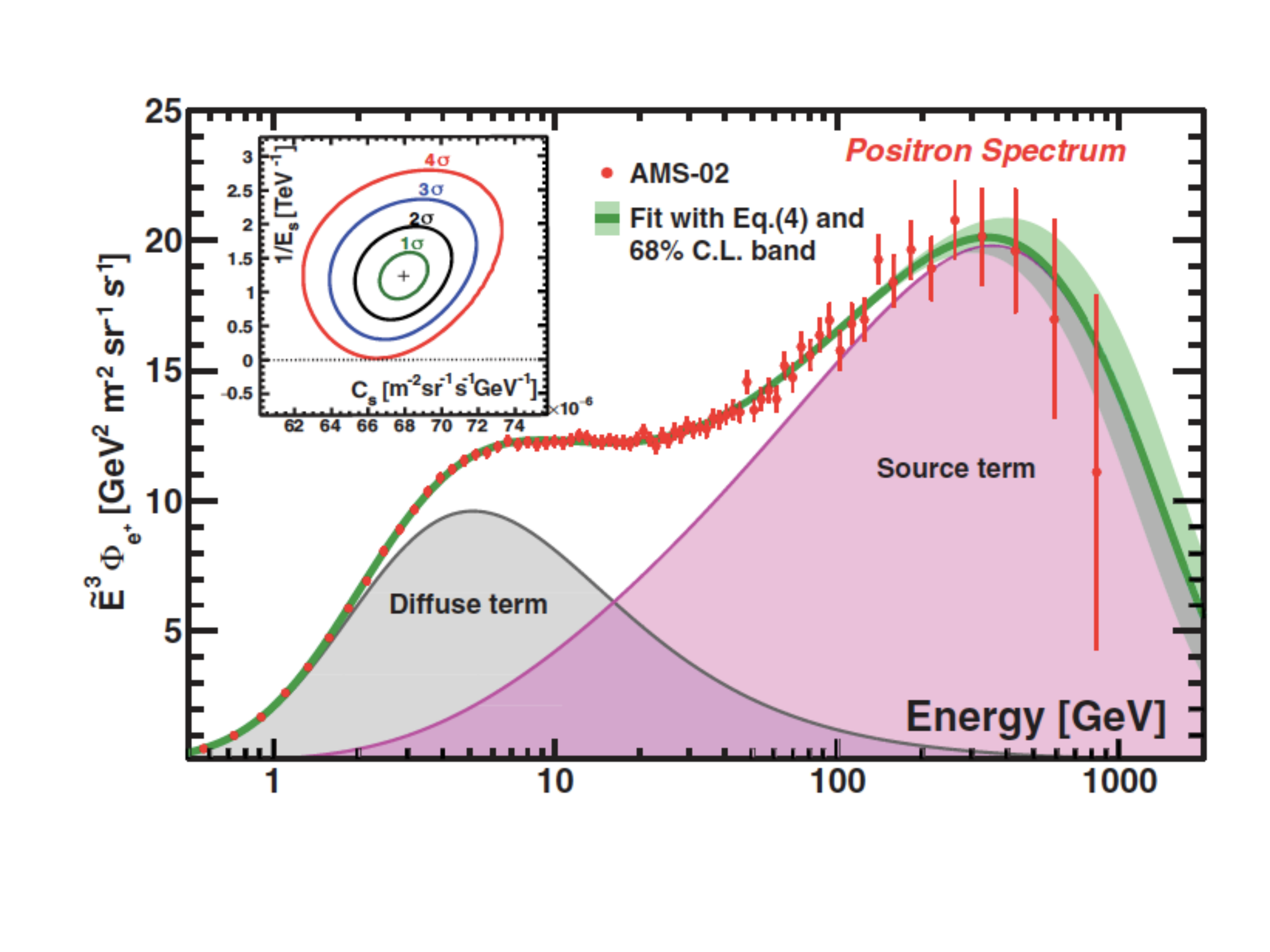,width=9cm}
\vspace{-1.3cm}
\caption{The AMS-02 positron spectrum \cite{AMSPos} and its ``diffuse" and ``source" terms.
The green line is their sum.}
\label{fig:Them}
\end{figure}

The flux $F(E_+)$ of cosmic-ray (CR)  positrons, measured by AMS,
is shown as $E_+^3\,F(E_+)$ in Fig.(\ref{fig:Them}), copied here from \cite{AMSPos}. In it the 
 ``Diffuse term" represents the contribution of CR protons and nuclei interacting with the matter of the
interstellar medium (ISM) to produce secondary positrons. Concerning this term, the cited authors  state:
{\it Explicitly, we have chosen the first term of Eq. (4) based
on the general trend of the commonly used cosmic ray
propagation models, even though all have large
uncertainties, but all show a maximum of the spectrum
below 10 GeV.} 

The ``source" in ``Source term" in Fig.(\ref{fig:Them})
refers to the contention that the positron excess originates
in the source of CRs. More or less conventional sources have been discussed by various authors
\cite{source}.

\begin{figure}[]
\vspace{-.5cm}
\hspace{-.5cm}
\centering
\epsfig{file=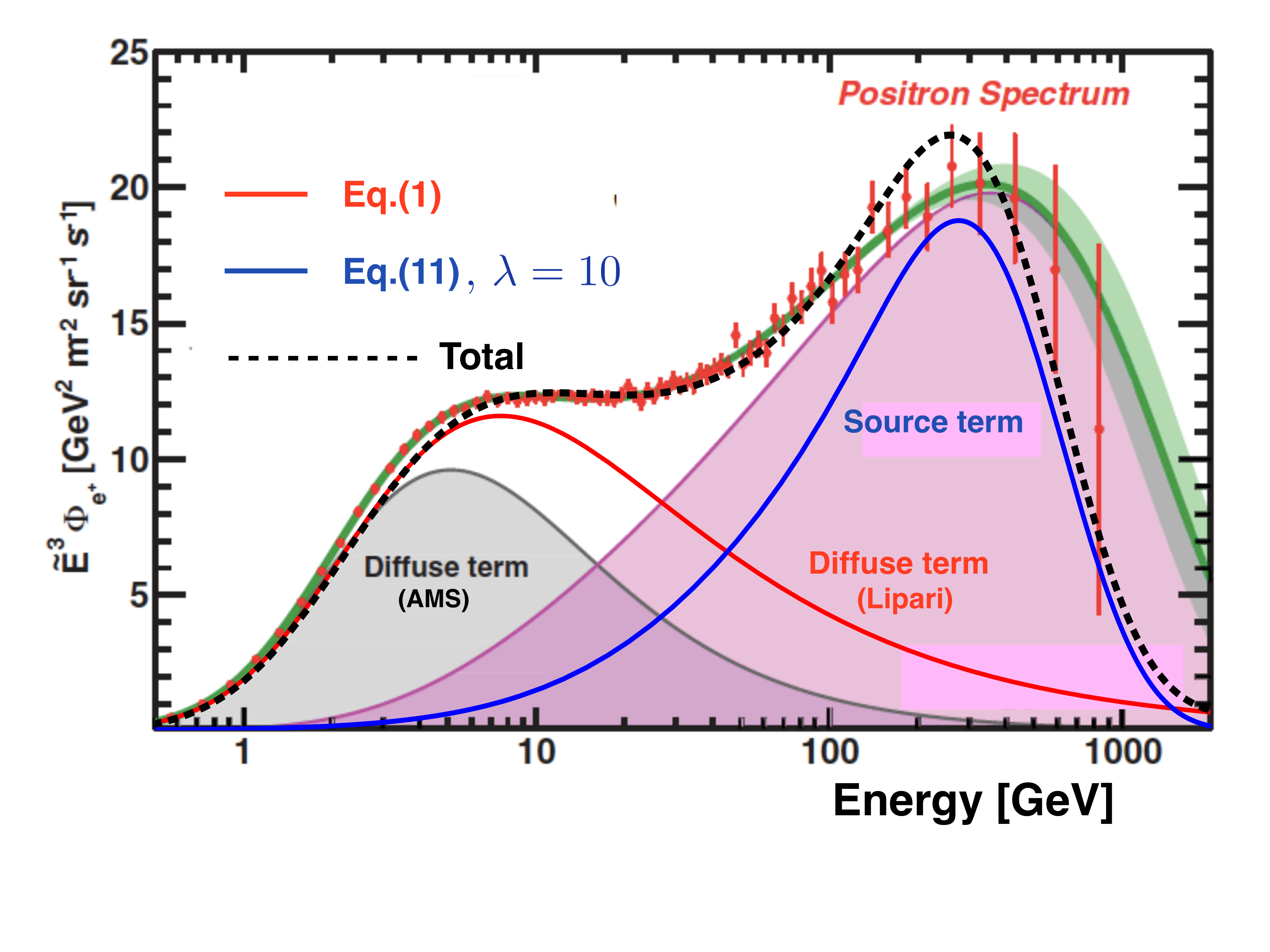,width=9cm}
\vspace{-1.3cm}
\caption{Our adopted diffuse term (red) and calculated source term (blue), and their sum
(black, dashed).}
\label{fig:Results}
\end{figure}

In discussing quantitatively the ``Source term" in Fig.(\ref{fig:Them}) it is important to choose 
the diffuse term
with heed. Lipari \cite{Lipari} has proposed a carefully built 
parametrization of the earlier  AMS data \cite{AMSold}. It has
two terms, the second of which (the would-be source term)
accounts for the observed change of spectral slope
at $E_+\simeq 20$ GeV. Without this contribution Lipari's fit is:
\begin{equation}
{d\Phi_{e^+}\over dE_+}= 
 K {E_+^2\over (E_+ + \epsilon)^2} \left[{E_+ + \epsilon\over E_0}\right]^{-\Gamma}
 \label{eq:Lipari}
 \end{equation}
 with $K=\rm 0.018/ (GeV\, m^2\, s\, sr)$, $\epsilon=0.94$ GeV, $E_0=10$ GeV and
 $\Gamma=3.62$. This will be the diffuse term adopted here. Multiplied
 by $E_+^3$, it is shown as the red line in 
 Fig.(\ref{fig:Results}).
The blue line in Fig.(\ref{fig:Results}) is the source term we shall derive. Summed to the diffuse term
it results in the black dotted line. The only arbitrarily chosen parameter in constructing this figure is the 
overall normalization of the source term. But, as we shall see, the calculated normalization is,
in spite of inevitable uncertainties, remarkably compatible with the observed one.

\section{The model}

More than a decade ago a Cannon-Ball (CB) model of Gamma Ray Bursts (GRBs) 
and X-Ray Flashes (XRFs) was
proposed \cite{DP} and elaborated \cite{DD}. The model is inspired by the observations of 
relativistic jets of matter emitted
by quasars, and by microquasars such as GRS 1915+105 \cite{GRS}.
A-periodically, about once a month, this black hole launches two
oppositely directed {\it cannonballs}, traveling at $v\sim 0.92\, c$.
When this happens, the continuous X-ray emissions
--attributed to an unstable accretion disk fed by a ``donor" star-- temporarily decrease.
 %The mechanism responsible for these ejections,
%due to episodes of violent accretion into a very
%massive black hole, is not understood in detail.
 
 The `cannon' of the CB model is analogous to the ones
of quasars and microquasars. In the {core-collapse} responsible for a stripped-envelope
SNIc event, due to the parent star's rotation, a short-lived accretion disk  is produced around
the newly-born compact object.
%,  by stellar material originally
%close to the surface of the imploding core, or by more distant stellar matter
%falling back after the shock's passage. 
 
 A CB made of {\it ordinary-matter plasma} is emitted, as
in microquasars, when part of the accretion disk
falls abruptly onto the compact object. {\it Long-duration} GRBs  
%and {\it non-solar} CRs 
are produced by these jetted CBs. 
The {\it `inverse' Compton scattering} (ICS) of ambient light by the electrons within a CB  
produces a highly forward-collimated beam of higher-energy photons.
Seen close to the CB's direction of motion, the beam of $\gamma$-rays is
a pulse of a GRB. Not so close, it is the pulse of an XRF. To agree with 
observations, CBs must be launched with typical Lorentz factors (LFs), 
$\gamma={\cal{O}} (10^3)$, and baryon numbers, 
$N_{\!B}={\cal{O}}(10^{50})\sim 10^{-7} N_\odot$ \cite{DD}.

The way the CB model describes the data and its many  
predictions regarding CRs and GRBs are
summarized in the appendix of \cite{ADR}. Comparisons with the ``standard
models" of GRBs are discussed in \cite{DDComp}.

The CB model is also a model of primary non-solar CRs \cite{DD2008}. 
In that article, we argued that the model offers a very good description of the spectra and 
abundances of CR nuclei, and of electrons, with several priors chosen in their
allowed ranges, but only one parameter to be fit to the data. The data have improved during the
last decade. For example,  measurements of the proton ``knee" have been refined and
knees in the He, Fe \cite{Hor} and $e^++e^-$ \cite{epluspexps}
spectra have been
observed, precisely as predicted by the CB model of CRs \cite{ADR}. In a nutshell:
CRs are made by CBs as they scatter the constituents of the ISM. The maximum energy
of a CR of mass $M$ is $2\,\gamma^2\,M$, the limit of forward scattering.
With the CBs' LFs, $\gamma$, distributed around the typical value of  ${\cal O}(10^3)$,
the correspondingly gradual cutoff  results in
the correct positions and shapes of the cited knees \cite{ADR}. 

\section{The CR $e^+$ spectrum}

\subsection{The ``wind"}
\label{ss:wind}

The $e^+$ source flux of Fig.(\ref{fig:Results}) is produced in the neighborhood of the
CB-launching supernova (SN), as it traverses the SN's close environment.
Massive stars lose mass in the form of ``winds", before they die in SN explosions. 
We shall refer to the pre-SN close-by material, accumulated by previous 
ejecta, as ``the wind", for short. As discussed in great detail in \cite{DD},
the relevant observations \cite{winds,winds2} indicate very high wind particle-number densities, 
$n\sim 5\times 10^{7}$ cm$^{-3}$, at the distances, $l= {\cal{O}}(10^{16})$
cm, of interest to the production of GRBs and positrons in the CB model. The measured $n$
and mass density,  $\rho$,
decline roughly as $1/l^2$ and the wind's  {\it ``surface density''} is 
$\Sigma\equiv \rho\, l^2 \sim 10^{16}\; {\rm g\; cm^{-1}}$ \cite{winds,winds2}.

The CB-model input priors are shown in Table I.

\begin{table}
      \caption[]{\bf Input priors of a CB and the SN's
      ``wind''.}
\vskip -0.2cm
            $$ 
            %\begin{array}{ p{0.13\linewidth}l p{0.28\linewidth}l}
             \begin{array}{ p{0.18\linewidth}l p{0.50\linewidth}l}
            \hline
            \noalign{\smallskip}
            Parameter     &  \rm Value & Definition \\
            \noalign{\smallskip}
            \hline
            \noalign{\smallskip}
 $\;\;\;\;\;\:\;\gamma$      & {\cal O}(10^3) & CB's Lorentz factor$^{\mathrm{a,}}$$^{\mathrm{b}}$ \\
% $\;\;\;\;\;\:\;\beta_s$ & \sim 1 & Initial expansion velocity\\
 $\;\;\;\;\;\:\; N_{\!B}$ & 10^{50} & CB's baryon  number$^{\mathrm{a}}$ \\
 $\;\;\;\;\;\:\; c_s$ & c/\sqrt{3}& CB's expansion velocity$^{\mathrm{a}}$ \\
  $\;\;\;\;\;\:\; \Sigma$ & 10^{16} {\rm g/cm}& Wind's surface density$^{\mathrm{a}}$ \\
\hline
         \end{array}
     $$ 
\begin{list}{}{}
\item[$^{\mathrm{a}}$] Typical CB-model value \cite{DD}.
\item[$^{\mathrm{b}}$] The $\gamma$ distribution is that of \cite{ADR}, here Eq.(\ref{eq:Dgamma}).
\end{list}
   \end{table}

\subsection{A CB sailing in the wind}

In what follows, to avoid pedantic factors close to unity, we consider the
composition of a CB (but not of the SN's wind) to be that of hydrogen. At 
$\gamma={\cal O}(10^3)$,
the $pp$ total cross section, $\sigma_{pp}\approx 40$ mb, is dominantly inelastic.
The $p$-nucleus cross section is an incoherent sum over the
$p$-nucleon cross sections, with $\sigma_{pn}\approx\sigma_{pp}$.
The CB's radius of collisional transparency to the ambient protons or nuclei
is $R_{pp}\!\sim \![3 \,\sigma_{pp}\,N_{\!B}/(4\pi)]^{1/2}$ $\sim\! 10^{12}$ cm,
for  $N_{\! B}\!=\! 10^{50}$. Since the CB's 
initial internal radiation pressure is large, it should expand (in its rest frame) at a radial velocity 
 $c_s\!=\!c/\sqrt{3}$, the speed of sound in a relativistic plasma.
When the $pp$ collisions cease the CB has travelled a distance 
$l_{max}(\gamma)\!=\!\sqrt{3}\,\gamma\,R_{pp}\!=\!1.7\times10^{15}\,(\gamma/10^3)\,{\rm cm}$,
where the LF reflects the relation between the times in the CB and SN rest systems.

What fraction of the CB's protons is lost to interactions with the wind?
Let the transverse radius of a CB at a distance $l$ from the SN be $r_{_{\!\rm CB}}(l)$, corresponding
to a surface $S(l)=\pi\,r_{_{\!\rm CB}}^2=\pi\,l^2/(3\,\gamma^2)$. The wind's baryon-number density 
is $n(l)\approx \Sigma/(m_p\,l^2)$, with $m_p$ the proton's  mass. The number of 
$pp$ plus $pn$ collisions ($p{\cal \small  N}$) is:
\begin{equation}
N_{p{\cal \small  N}}\!=\!\int_0^{l_{max}(\gamma)} S\,n\, dl={\pi\,R_{pp}\over{\sqrt{3}\,\gamma}}\,{\Sigma\over m_p}
=1.1\times 10^{49} \,{10^3\over\gamma},
\label{Npp}
\end{equation}
so that a CB with $\gamma=10^3$ would have lost
$\sim 10\%$ of its $\sim 10^{50}$ protons to $p$-wind interactions. Kinematically, a negligible effect.

\subsection{Surviving attenuation by the wind}
\label{ss:attenuation}

The $p$-wind collisions give rise, mainly via the chain 
$pp\,({\rm or} \, pn)\!\to\! \pi\,{\rm or}\, K\!\to\! \mu^+\!\to\! e^+$, to the source positrons of interest here.
Not all positrons, however, manage to penetrate the SN's wind environment.
Let $\sigma_T=0.665\times 10^{-24}$ cm$^{-2}$ be the Thomson cross section, adequate for the
study, quite independently of the nuclear target, of the penetrability of the wind to the $\gamma$-rays
of a GRB \cite{DD}. Here we are interested in $\gamma$-rays or positrons of much higher energy,
whose attenuation lengths by a pure-element intervening material are very similar, 
but decrease by an order of magnitude
from H to Pb. We do not know the precise composition of the wind. In the wind of e.g.~SN1997eg, 
H, He, N, O, Mg, Si and Fe lines have been observed \cite{winds2}. 
We shall adopt, as a compromise or average, 
the $\gamma$-O or $e^+$-O attenuation 
cross section which, above an energy of $\sim 1$ GeV, is very close 
to constant and to $\sigma_T$ \cite{attenuation}.

Not all the positrons made in $p$-wind collisions escape
unscathed to become observable: they may be reabsorbed by the wind's
material. The probability that an $e^+$ produced at a distance $l$ from the SN
evades this fate, in a wind with a density profile $n_e\propto l^{-2}$,
is $A(l)= {\rm exp}[-(l^w_{tr}/l)^2]$ with
$l^w_{tr}=\sigma_T\, \Sigma/m_p$  the distance at which the 
remaining ``optical" depth of the wind is unity. 

Still referring to a single CB with $N_{\! B}=10^{50}$ and LF $\gamma$, let us
estimate the number, $N_{\to e}$, of proton-wind collisions whose produced positrons
penetrate the wind unscathed. To do so, add an extra factor $A(l)$ to the integrand in 
Eq.(\ref{Npp})... and integrate. The result is:
\begin{eqnarray}
&&N_{\to e}(\gamma)={\pi\,R_{pp}\over{3\,\gamma^2}}\,{\Sigma\over m_p}\,I[l^w_{tr},l_{max}(\gamma)],
\nonumber\\
I&\equiv& l_{max}\,{\rm exp}[- (l^w_{tr}/l_{max})^2]-\sqrt{\pi}\,{\rm erfc}[l^w_{tr}/l_{max}].
\label{Nppe}
\end{eqnarray}

\subsection{The $e^+$ energy distribution}

Let $F(x,E_p)$, with $x\equiv E_+/E_p$, be the $x$ distribution of positrons normalized 
(i.e.~$x$-integrated) to their multiplicity at $E_p=\gamma\,m_p$. We shall use the $F(x,E_p)$
calculated in \cite{MosStr} and \cite{KAB}, which agree at the relevant $E_p\!=\!{\cal{O}}\,\rm (TeVs)$,
and are easy to check, since at such high energies the $pp\,{\rm  or}\,pn\!\to\! K,\pi$ yields approximately scale and all particles in the decay chain to positrons are approximately collinear. The simplest input to use
are Eqs.(62 to 65) of \cite{KAB}, corrected by an --admittedly cosmetic--
factor $\sim\! 5/6$ for the observed and expected primary
charged-particle multiplicities in $pp\,{\rm  and}\,pn$ collisions \cite{multip}.

The number distribution of positrons as a function of their energy,
once more referring to a single CB with $N_{\! B}=10^{50}$ and LF $\gamma$, is:
\begin{eqnarray}
{dn(\gamma)\over dE_+}&=&\int_0^1 \!dx\, F[x,m_p\,\gamma]\,N_{\to e}(\gamma)\,
\delta(E_+ - m_p\,\gamma\,x) \nonumber\\
 &=& {1\over m_p\,\gamma} \,F\!\left[{E_+\over {m_p\,\gamma}},m_p\,\gamma\right]\,N_{\to e}(\gamma).
 \label{dndE}
\end{eqnarray}

%The number distribution of positrons of energy $E_+$ made by a CB of incident LF $\gamma$ is:
%\begin{equation} 
%{dN(\gamma)\over dE_+}={1\over m_p\,\gamma} 
%F\left[ {E_+\over{m_p\,\gamma}},\gamma\,m_p\right] N_{\to e}(\gamma).
%\end{equation} 

\subsection{The $\gamma$ distribution of CBs}
\label{ss:CBgamma}

Next, we ought to weigh the above result with the distribution, $D(\gamma)$, of the Lorentz factors of 
CBs. We adopt the one that describes the CR proton knee
and was used to predict \cite{ADR} the shapes and positions of the He, Fe \cite{Hor} and 
$e^++e^-$ \cite{epluspexps} knees. To wit:
\begin{eqnarray}
&&D(y)={\rm exp}(-[(y-y_0)/\sigma]^2);\nonumber\\
&&y\equiv {\rm Log}_{10}[\gamma^2],\;\;y_0=6.3,\;\;\sigma=0.5.
\label{eq:Dgamma}
\end{eqnarray}

The energy distribution of positrons made by a CB of $N_B=10^{50}$,
whose LF is randomly picked from the above distribution of CB's LFs is, {\it at the source}:
\begin{equation}
{dN\over dE_+}=\int \widetilde D(\gamma)\,{dn(\gamma)\over dE_+}\,d\gamma,
\label{dNdE2}
\end{equation} 
with $dn(\gamma)/ dE_+$ as in Eq.(\ref{dndE}) and $\widetilde D\equiv D/\int D\,d\gamma$. 

\begin{figure}[]
\vspace{.2cm}
\hspace{-.5cm}
\centering
\epsfig{file=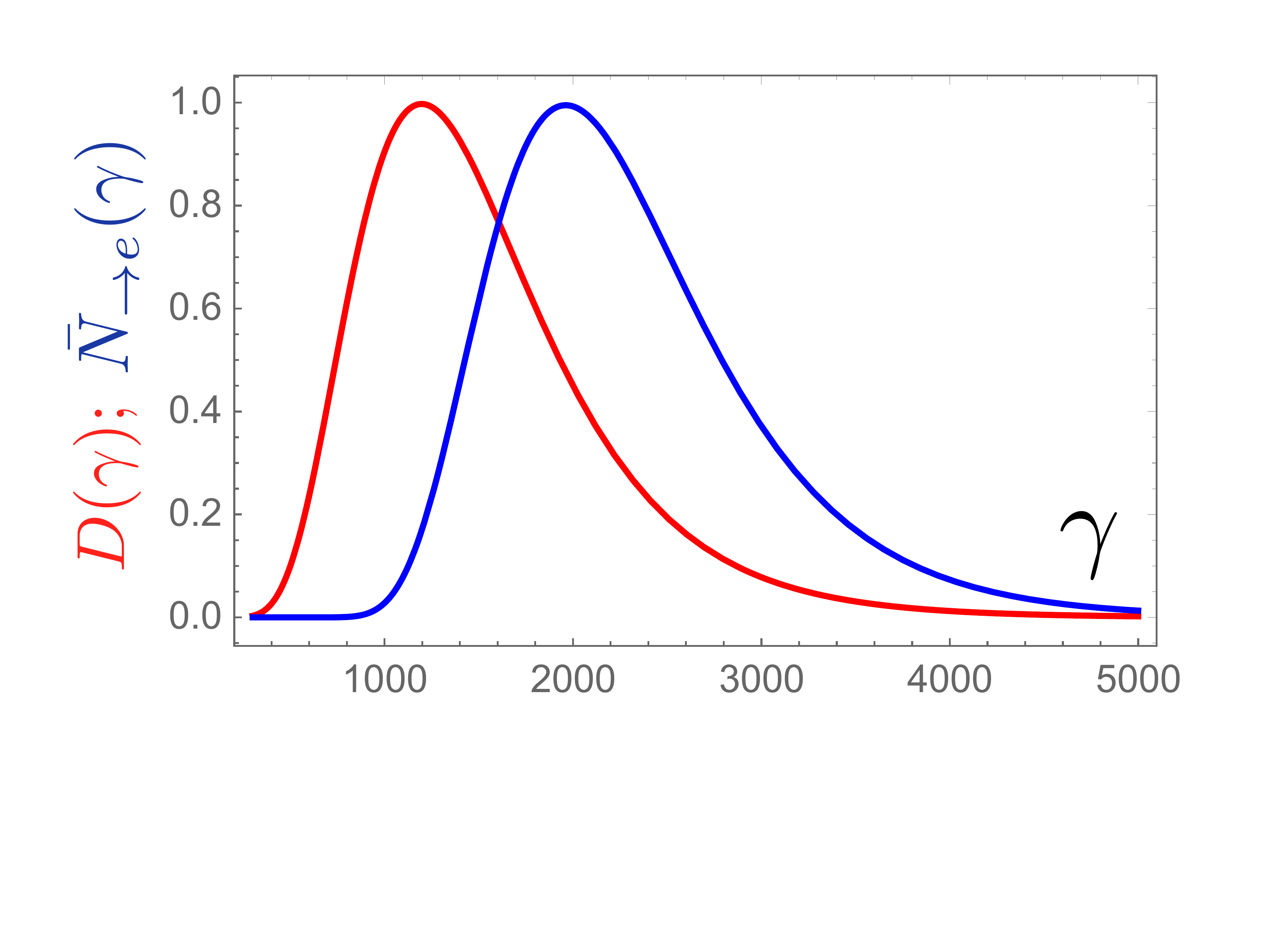,width=9cm}
\vspace{-2cm}
\caption{In red $D(\gamma)$. In blue $\bar N_{\to e}(\gamma)$. Both arbitrarily
normalized in the figure.}
\label{fig:GammaDistrs}
\end{figure}

Incidentally, the function 
$\bar N_{\to e}=\widetilde D \,N_{\to e}$ is the number distribution of proton-wind collisions 
resulting in positrons that penetrate the wind, for a single CB of $N_B=10^{50}$, and $\gamma$
randomly picked from the distribution $\widetilde D$. The shapes of $\widetilde D(\gamma)$ and
$\bar N_{\to e}(\gamma)$ are drawn in Fig.(\ref{fig:GammaDistrs}), showing how
$\bar N_{\to e}(\gamma)$ is weighed to higher LFs than $D(\gamma)$ because larger-$\gamma$ CBs 
keep interacting with the wind up to distances at which the latter is getting thinner.

\subsection{The galactic rate of CB-emitting supernovae}
\label{ss:SNe}

Comparing the observed SN and 
long-duration GRB rates we concluded in \cite{DD} that {\it the GRB rate is
consistent with being equal to the total rate of core-collapse SNe, or to a fraction of it
that may be as small as $\sim\! 1/4$}. 
It has been recently shown
that at small redshifts, $z \!<\! 0.15$, the fraction of long duration GRBs without an associated bright 
SN is comparable to that of GRBs associated with SNe \cite{DDz}.
Crucial to these conclusions is the fact that in the CB model
only a few per million GRBs are observable, since their ICS radiation is beamed 
within a cone of opening angle $\theta\!\sim\! 1/\gamma\!\sim\! 1$ mrad. 
Short GRBs, also emitting CBs in binary neutron star mergers \cite{DDDSHB},
 are only a fraction $\sim\! 7.35$\% of the long ones \cite{Grenier}. Finally,
a recent study confirms that the SN rate in the Galaxy is about two per century
\cite{Diehl}. 
 
Abridging the previous paragraph into one number, we
 shall normalize the coming results to $N^{\rm SN}_{ \rm Gal}=1$, that is:
one GRB-generating galactic SN per century. The observed median number of clear peaks
(i.e.~CBs) in long-duration GRB light curves is $\sim 5$ \cite{GRBpeaks}.
Since only ``one-side" of a GRB is observable we set $N_{\rm CB}=10$, so that
Eq.(\ref{dNdE2}) must be multiplied by this number for one GRB-generating galactic SN.

\section{$e^+$ escape and energy losses}
\label{s:out}

Next we must take into account how the $e^+$
 source spectrum of Eq.(\ref{dNdE2}) is affected by energy losses
in their interactions with the radiation and magnetic fields of the Galaxy, as well as their possible
escape therefrom. 

\subsection{Positron escape}
\label{ss:escape}

Let us first discuss the escape from the Galaxy, characterized by a confinement time:
\begin{eqnarray}
\!\!\tau_{\rm conf}({E_+})=\tau_0\,
(1\,{\rm GeV}/{E_+})^{\beta_{\rm conf}};\nonumber\\
\tau_0\sim 2.5 \times 10^7\,{\rm years},\;{\beta_{\rm conf}}\sim 0.6,
%\pm 0.1,
\label{eq:conf}
\end{eqnarray}
with $\tau_0$ and ${\beta_{\rm conf}}$ estimated from
observations of astrophysical and solar plasmas and corroborated
by measurements of the relative abundances of secondary CR isotopes  \cite{CONFI}.
Recent measurements of the B/C ratio \cite{AMS} imply, at low rigidity, $R$, smaller 
values of $\beta_{\rm conf}$ than given in Eq.(\ref{eq:conf}), see Fig.(2) in \cite{AMS}. But,
as theoretically expected, $\beta_{\rm conf}(R)$ flattens as $R$ increases.
The measured value at the highest-rigidity point ($R=860$ GV) is $.52\pm.13$. 
The rigidities of the positrons discussed here are similar.
It is therefore reasonable, as we do in what follows, to adopt the ``traditional"
$\beta_{\rm conf}=0.6$, compatible with the results of \cite{CONFI} and \cite{AMS}.
The outcome for a more ``theoretical" choice at relatively low $E_+$, 
$\beta_{\rm conf}=1/3$ \cite{Kolmo} is not significantly different.

\subsection{Positron energy loss by ICS of starlight}
\label{ss:light}

The $e^+$ energy loss of positrons due to ICS on ambient light in the near-UV to near-IR
regime requires a detailed treatment \cite{DDeloss}, if only because it is
%, as we shall see,
the only energy-loss mechanism that might significantly affect the {\it shape} of the observed
$e^+$ spectrum. 

The energy density of the interstellar radiation field of
the Galaxy, $U$, has been carefully modeled in \cite{PS,PMS}, allowing one to
compare its local value, $U_{\rm loc}$, some 7.5 kpc away from the galactic center,
to its values, $U_{\rm in}$, in the inner disk at 0 to 3 kpc from the center, the domain
where most galactic SNe occur. At wavelengths from 0.1 to 10 $\mu$m --to
which we shall refer to as ``starlight"-- the 
values of $U_{\rm in}$ are $\sim\!5$ to $\sim\!30$ 
times larger than $U_{\rm loc}$. In this wave-lenght domain
we shall adopt the value $U_{\rm in}= 10\,U_{\rm loc}$, but compare the results
with the ones for $U_{\rm in}= U_{\rm loc}\sim 0.39$ eV/cm$^3$, to illustrate the sensitivity to this input.
The values in the Far InfraRed (FIR) domain (wavelengths from 15 to
$10^3$ $\mu$m) the difference between $U_{\rm in}$ and  $U_{\rm loc}$
is  less pronounced \cite{PS,PMS}.

The ICS of starlight by positrons in the energy domain we study brackets the
transition from a Thomson to a Klein-Nishina cross section. In this respect we
follow the analysis in \cite{SREloss} and \cite{DDeloss}. Define an energy-loss time, $\tau_\star$:
\begin{eqnarray}
\tau_\star(E_+,\lambda) &=& 3 \,m_e^2/[4\, \sigma_T\, c\,E_+\, U_\star (E_+,\lambda) ] ,
\nonumber\\
U_\star (E_+,\lambda) &=& \lambda\, U_{\rm loc} \, E_{\rm KN}^2/(E_{\rm KN}^2 + E_+^2) ,
\nonumber\\
E_{\rm KN}&=&0.27\,(m_e\,c^2)^2/{(k\,T)},
\label{eq:KN}
\end{eqnarray}
where $\lambda$ is introduced to study varying the amount of starlight energy density
in the inner Galaxy, and
$E_{\rm KN} = 140$ GeV for light at a temperature $T=5700$ K \cite{SREloss}.

\subsection{Other $e^+$ energy losses and the combined lifetime}
\label{ss:other}

Three other sources affecting the local spectrum of positrons are their 
interactions with ambient photons of various types: 
the real ones of the Cosmic Background Radiation (CBR) and the FIR, 
and the virtual ones of magnetic fields (B) --ICS on the latter photons is usually called
bremsstrahlung.
Choose these photon energy-densities to be the locally measured
ones $(U_{\rm B},\,U_{\rm FIR},\,U_{\rm CBR}) \!\sim\! (0.4,\,0.4,\,0.26)$ eV/cm$^3$
and define $U_{\rm soft}=U_{\rm B}+U_{\rm FIR}+U_{\rm CBR}$.
The corresponding positron lifetime is:
\begin{equation}
\tau_{\rm soft}(E_+)= 3 \,m_e^2/[4\, \sigma_T\, c\,E_+\, U_{\rm soft} ].
\label{eq:soft}
\end{equation}

\begin{figure}[]
%\vspace{-.5cm}
\hspace{-1.cm}
\centering
\epsfig{file=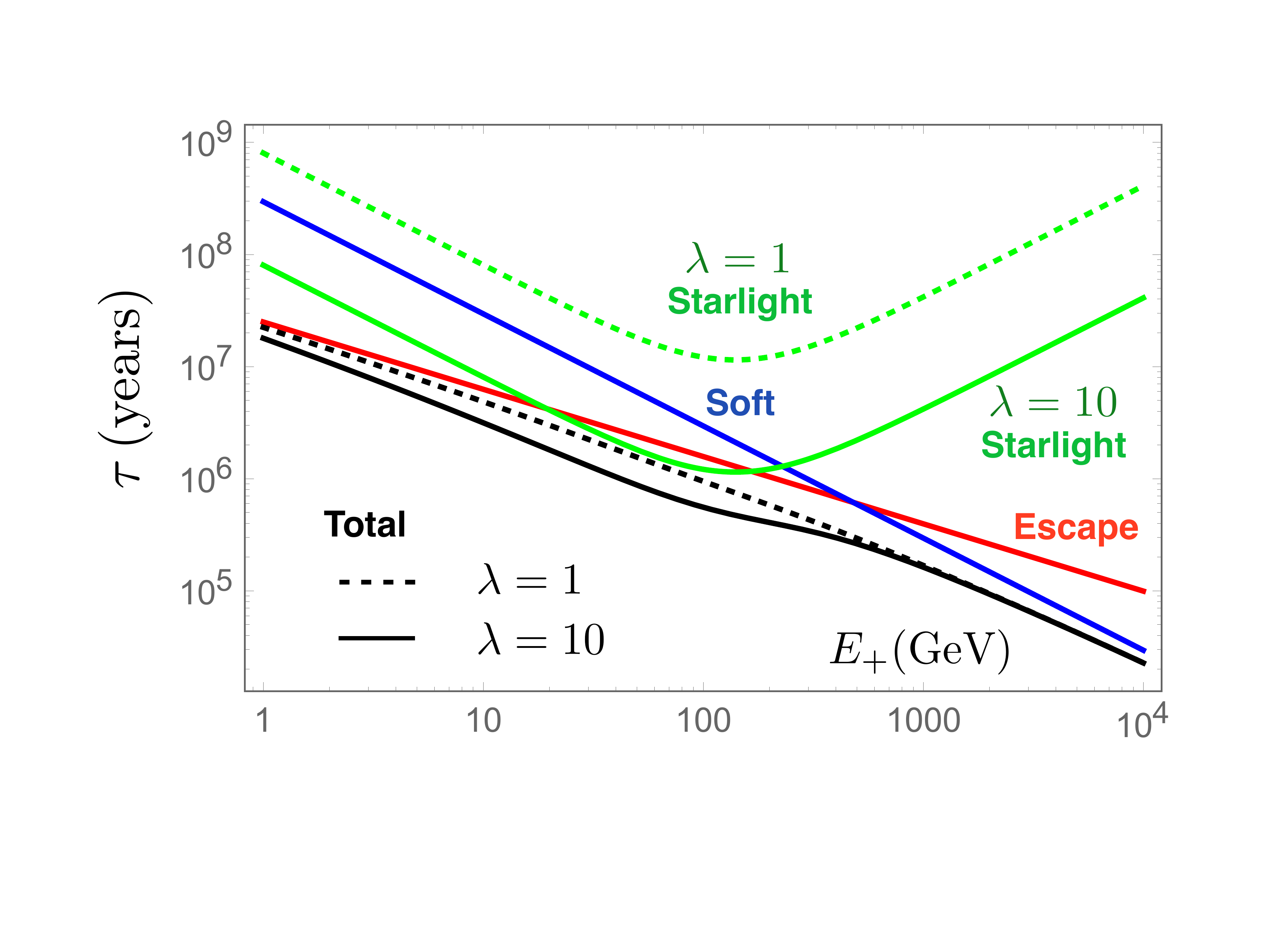,width=9.5cm}
\vspace{-1.5cm}
\caption{Positron lifetimes as functions of energy. The red ``Escape" line is Eq.(\ref{eq:conf}).
The green lines are for the effect of starlight and two values of $\lambda$ in Eq.(\ref{eq:KN}). 
The blue ``Soft" line is Eq.(\ref{eq:soft}). The black ``Total" lines are for two values of $\lambda$
 in Eq.(\ref{eq:tautot}).}
\label{fig:tau}
\end{figure}

To obtain the positron lifetime, $\tau$, combining the effects of escape and the various
energy-loss mechanisms, one inverts the sum of the inverse separate lifetimes:
\begin{equation}
\tau(E_+,\lambda)={1\over \tau^{-1}_{\rm conf}({E_+})+\tau^{-1}_\star(E_+,\lambda)
+ \tau^{-1}_{\rm soft}(E_+)} \, .
\label{eq:tautot}
\end{equation}
The various lifetimes we have discussed are shown in Fig.(\ref{fig:tau}). Notice 
the shapes of $\tau(E_+,1)$ and $\tau(E_+,10)$.

\section{The local flux of positrons}

To recapitulate and proceed, the number flux of Eq.(\ref{dNdE2}), multiplied by $N_{\rm CB}\!=\!10$
(the median number of CBs), by $N^{\rm SN}_{ \rm Gal}\!=\!1/$century
 (the galactic rate of CB-emitting SNe), and
by the lifetime $\tau(E_+,\lambda)$ of Eq.(\ref{eq:tautot}) (expressed in centuries) is the 
predicted number flux
per unit energy interval of the source positrons that will still be in the Galaxy and have an
energy $E_+$. To convert this result into a flux per stereo-radian and per unit surface, we must
still multiply it by $c/(4\, \pi)$ and distribute the positrons over the Galaxy. One way of doing the latter
is to use a state-of-the-art CR propagation code, such as
GALPROP \cite{GALPROP0,GALPROP} or DRAGON \cite{DRAGON}, in one of their
many options. Instead we use a simpler procedure that conveys the gist of the argument.
We do this in two steps:

In a first very rough estimate, assume the positrons to be uniformly distributed in a 
``leaky box" cylinder with the usually assumed
dimensions: 15 kpc radius and 4 kpc (half-)hight, that is $V\!=\!1.66\times 10^{68}$ cm$^3$.
The result for the $e^+$ flux $F$ is then:
\begin{equation} 
{d\Phi_{e^+}(\lambda)\over dE_+}=N^{\rm SN}_{ \rm Gal}\,N_{\rm CB}{dN\over dE_+}
\tau(E_+,\lambda){c\over {4\,\pi\,V}}\, ,
\label{eq:result}
\end{equation}

Second, we reinterpret this outcome by assuming that at our position in the Galaxy (half-way
to the rim of the leaky box) the number density of positrons, which ought to decrease
with distance to the galactic centre, is close to the above 
average. State-of-the art calculations not based on a leaky box support this claim.
For instance, Fig.(19) of \cite{StMos} or Fig.(6.8) of \cite{DGagg}. 
The latter figure is also for CR protons, not positrons,
but is computed with ``a Plain Diffusion setup" with DRAGON \cite{DRAGON} (protons and positrons
lose energy at different rates, but at relativistic energies they diffuse in the same manner).

The result of Eq.(\ref{eq:result}), with $\lambda=10$, and duly multiplied by $E_+^3$, is the blue 
``source" term in Fig.(\ref{fig:Results})... but for the fact that the predicted normalization is 
1.23 times larger.

The above result on the normalization is either an incredible coincidence or a very satisfactory
consistency check, given the admittedly large number of accumulated uncertainties 
--a non-obvious (i.e., non-linear) example:
an increase by 40\% (or 19\%) of the assumed wind surface density, $\Sigma$, would reduce the
 cited normalization by a factor of two (or 1/1.23).  

\section{Minor points of discussion}
 
 Regarding the shape of our source term, various comments are illustrated in Fig.(\ref{fig:Try}).
 There, the blue curve is Eq.(\ref{eq:result}) with $\lambda\!=\!1$, that is with the 
 starlight energy density measured in the solar neighborhood. It is to be compared with the blue 
 ($\lambda\!=\!10$) result of Fig.(\ref{fig:Results}), for which the Thomson to Klein-Nishina
 transition is more pronounced. The red curve in Fig.(\ref{fig:Try}) has $\lambda\!=\!20$ and
 a distribution of LFs, Eq.({\ref{eq:Dgamma}), centered at the same maximum, but slightly broader
 ($y_0\!=\!6.53$, $\sigma\!=\!0.8$). A LF distribution enhanced a large $\gamma$ in Eq.(\ref{dNdE2})
 is a way to compensate for the fact that the $F(x,E_p)$ inputs we have used \cite{MosStr,KAB}
 lack positrons made in the decay chains
 of charmed particles. These positrons are relatively few, but more
 energetic than the ones from $\pi$ and $K$ decays.

\begin{figure}[]
\vspace{.2cm}
\hspace{-.5cm}
\centering
\epsfig{file=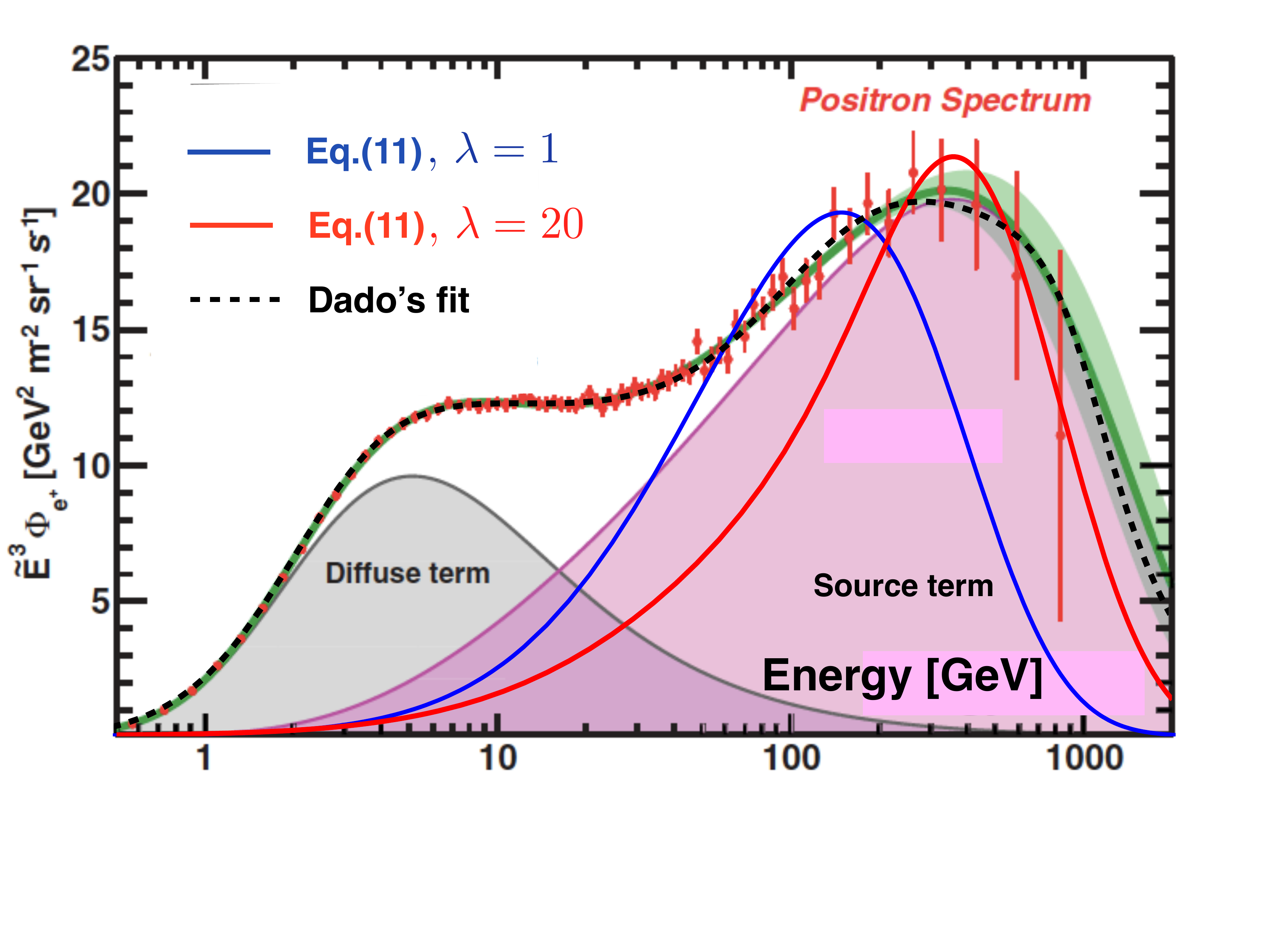,width=9cm}
\vspace{-1.5cm}
\caption{Sensitivity to various values of the parameters.}
%\vspace{-0.9cm}
\label{fig:Try}
\end{figure}

It should be clear that one could obtain a perfect description of the data by fitting various parameters,
such as $\lambda$ and the energy density of magnetic fields and the FIR. This was done,
with otherwise much less detail, in \cite{DDAMS}. The black dashed curve in Fig.(\ref{fig:Try}) 
is a fit (Shlomo Dado, private communication) with $\chi^2/(\rm dof)\!=\!0.58$ and 9
fitted parameters, including the ones describing the diffuse term. Providing a perfect fit
may be esthetically satisfactory, but is not necessarily decisive since, after all,
there are no undebatable results for the diffuse term's size under the source term peak.

In the CB model the photons of a GRB pulse are made by
ICS of light by a CB's electrons and their typical number, integrated over energy and angles, is 
$N^\gamma_{\rm GRB} \!=\! 5\!\times\! 10^{52}$ \cite{DD}.
Our source positrons are accompanied by $\gamma$-rays in slightly higher numbers
and with a slightly harder spectrum \cite{MosStr, KAB}, forward collimated
within an angle $\sim p_{_{\rm T}}/{E}$ close to the GRB photon's opening angle,
$1/\gamma$.
In an observed GRB,  are these $\gamma$-rays
of higher $E$ than the typical $E_{\rm GRB}\!=\!250$ keV observable?

Alas, the answer
is so negative that a rough estimate suffices.
The ratio of $\gamma$-rays or $e^+$'s surviving wind absorption to the number of $pp$ or $pn$
collisions is $I/(\sqrt{3}\,\gamma)$, where Eqs.(\ref{Npp},\ref{Nppe}) have been used. 
At the maximum
of $\bar N_{\to e}(\gamma)$ at $\gamma\!\sim\! 2000$ in Fig.(\ref{fig:GammaDistrs}) 
this ratio is $\sim\!5\%$ of the number of $pp$ or $pn$ collisions at that same LF $\gamma$:
$5 \times 10^{48}$ according to Eq.(\ref{Npp}). All in all the number of prompt hard
photons per GRB pulse is $\sim 2.5\times 10^{47}$, that is a miserable
$5\times10^{-6}$ of $N^\gamma_{\rm GRB}$.
Moreover their spectrum, akin to the $e^+$ one of Eq.(\ref{dndE}),
is strongly peaked at low energies, with only a small fraction above, say, 1 GeV.

\section{Conclusion}
The CB model predicts the shape and normalization of the AMS positron spectrum
in a very satisfactory way.

\vspace{.6 cm}
\noindent {\bf Acknowledgment:} A. De R\'ujula acknowledges 
that this project has received funding/support from the European Union's Horizon 2020 research and innovation programme under the Marie Sklodowska-Curie grant agreement No 690575. I am particularly indebted to Shlomo Dado and Arnon Dar for discussions and advice.

%\vspace{.4 cm}

\end{document}